# COB-2023-0802
# ASSESMENT OF LARGE EDDY SIMULATION (LES) SUB-GRID SCALE MODELS ACCOUNTING FOR COMPRESSIBLE HOMOGENEOUS ISOTROPIC TURBULENCE

**Jhon Cordova, Cesar Celis**
Mechanical Engineering Section, Pontificia Universidad Católica del Perú
Av. Universitaria 1801, San Miguel, 15088, Lima, Perú
jcordovaa@pucp.edu.pe, ccelis@pucp.edu.pe

**Andres Mendiburu**
International Research Group for Energy Sustainability (IRGES)
Universidade Federal do Rio Grande do Sul
Rua Sarmento Leite 425, Porto Alegre, RS, Brazil
andresmendiburu@ufrgs.br

**Luis Bravo**
DEVCOM – US Army Research Laboratory
Aberdeen Proving Ground, MD, 21005
luis.g.bravorobles.civ@army.mil

**Prashant Khare**
Deparment of Aerospace Engineering
University of Cincinnati
Cincinnati, OH 45221-0070
Prashant.Khare@uc.edu

*Abstract. Most sub-grid scale (SGS) models employed in LES (large eddy simulation) formulations were originally developed for incompressible, single phase, inert flows and assume transfer of energy based on the classical energy cascade mechanism. Although they have been extended to numerically study compressible and reactive flows involving deflagrations and detonations, their accuracy in such sensitive and challenging flows is an open question. Therefore, there is a need for both assessing these existing SGS models and identifying the opportunities for proposing new ones, which properly characterize reacting flows in complex engine configurations such as those characterizing rotating detonation engines (RDEs). Accordingly, accounting for the decay of free homogeneous isotropic turbulence (HIT), this work provides a comparison of four different SGS models when compressibility effects are present, (i) the classical Smagorinsky model, (ii) the dynamic Smagorinsky model, (iii) the wall-adapting local eddy-viscosity (WALE) model, and (iv) the Vreman model. More specifically, SGS models are firstly implemented in the open-source computational tool PeleC, which is a high-fidelity finite-volume solver for compressible flows, and then numerical simulations are carried out using them. The second-order explicit method of lines (MOL) and the hybrid PPM/WENO-Z method are utilized here as temporal and spatial numerical schemes, respectively. The numerical simulations are initialized at a turbulent Mach number of $M_t = 0.6$ and a Taylor Reynolds number of $Re_\psi = 100$, which are two nondimensional parameters defining the initial kinetic energy in the domain, after which the energy is transferred from large to small scales until it is eventually dissipated by viscous effects. In terms of results, turbulent spectra, and the decay of physical quantities such as kinetic energy, enstrophy, temperature, and dilatation are computed for each SGS LES model and compared with direct numerical simulations (DNS) results available in literature. The LES numerical results obtained here highlight that the studied SGS models are capable of capturing the overall trends of all physical quantities accounted for. However, they also emphasize the need of improved SGS models capable of adequately describing turbulence dynamics in compressible flows. The SGS model comparisons discussed in this work will serve as a basis for proposing in the future new SGS model constructs focused on compressible reactive flows.*

## 1. INTRODUCTION

Large eddy simulation (LES) is a computational fluid dynamics (CFD) approach that has become increasingly popular in recent years for the simulation of turbulent flows. This is due to its turbulence scale separation process in which large energy-containing eddies are directly resolved, whereas small dissipative eddies are modeled using a sub-grid scale (SGS) model. In contrast to the turbulence scale treatment carried out in LES, Reynolds-averaged Navier-Stokes (RANS) simulations model all turbulent scales, whereas direct numerical simulations (DNS) resolve all turbulent scales without any modeling. Therefore, accounting for the capacity for resolving turbulent scales, DNS is the most accurate approach but it is still computationally expensive for flow configurations of practical interest (Berselli et al., 2007), whereas RANS is computationally efficient but makes significant approximations (Versteeg and Malalasekera, 2007). In this context, LES is a promising approach that provides higher accuracy than RANS while being less computationally expensive than DNS (Berselli et al., 2006). Notice however that the majority of studies on SGS modeling carried out in the past have been limited to incompressible flows, with some extensions including formulations for variable-density flows and low Mach number reactive flows (Poinsot and Veynante, 2005; Garnier et al., 2009; Gatski and Bonnet, 2013). In most cases, existing SGS models have been extended to compressible and reactive flows (Fureby, 2020) without major modifications. Therefore, the accuracy of current SGS models when used in compressible and reactive flows, in particular, in those sensitive and challenging ones involving deflagrations and detonations for instance, is still unclear (Garnier et al., 2009).

Rotating detonation engines (RDEs) represent a promising technology for propulsion systems that offers the potential for significant improvements in efficiency and power (Shaw et al., 2019). In short, the key to RDEs relies mainly on one or more detonation waves that rotate around the combustion chamber providing continuous thrust (Xie et al., 2020). Despite their several advantages, RDEs are complex systems involving multiple physical phenomena, such as combustion, turbulence, compressibility, and shock wave propagation (Fureby, 2020), which makes their modeling a challenging task. When modeling RDEs using LES, one of the most critical aspects involves properly capturing shock-turbulence-chemistry interactions (Fureby, 2022). From the compressible side, SGS models must account for the energy transferred between turbulent scales, which leads to the energy cascade behavior, and between modes, e.g., vortical energy can be transformed into acoustic energy or heat (Garnier et al., 2009). In this context, SGS models can significantly affect the accuracy and reliability of numerical results. Thus, following a strategy of gradually increasing geometrical and physical complexities, the assessment of current SGS models to properly characterize shock-turbulence interactions in unconfined systems is first required.

One of the most fundamental canonical cases involving shock-turbulence interactions is the one where a shock interacts with homogeneous isotropic turbulence (HIT). In this flow configuration the periodic boundary conditions employed in all three spatial directions remove the complexity of reproducing the near-wall turbulent behavior with the SGS model. For the evaluation of LES SGS models, this benchmark case can be of two types, (i) the sustained version, in which flow kinetic energy is maintained by injecting energy, and (ii) its counterpart, the freely decaying version, in which kinetic energy is eventually dissipated. In the decay of free HIT, specific aspects of turbulence dynamics can be assessed, including the energy spectra or the decay of physical quantities such as kinetic energy, enstrophy, temperature, and dilatation. Despite the fact of being a relatively simple case, few studies have been performed in the past considering compressible flows. For instance, in the DNS work of Motheau and Wakefield (2021 and 2020), the influence of different numerical schemes on turbulent compressible flow related parameters were evaluated. Similarly, using DNS, shock-turbulence interactions were analyzed by Wang et al. (2017). In addition, Wang et al. (2018) provided information about the inter-scale energy transfer when using Gaussian and spectral LES filters.

Accordingly, the present work aims to extend previous studies in the field of compressible turbulence. More specifically, classical SGS models developed in incompressible contexts are assessed accounting for the popular test case of free decay of HIT when compressible effects are present. The numerical simulations are carried out using the open-source computational tool PeleC (PeleC, 2023) considering four different SGS models, (i) Smagorinsky, (ii) dynamic Smagorinsky, (iii) wall-adapting local eddy-viscosity (WALE), and (iv) Vreman. Notice that, due to the similarities in their construction, the WALE and Vreman models were implemented into the PeleC following the rationale behind the implementation of the Smagorinsky model originally available in this tool. For validation purposes, DNS results of free decay of HIT available in the of Motheau and Wakedfield (2020) are utilized. The remainder of this document is organized as follows. Section 2 gives a brief overview of the governing equations and the closure models used here. Section 3 provides in turn a summary of the numerical methods employed and a short description of the HIT case studied in this work. The comparisons between the numerical results obtained here with the different SGS models accounted for and the DNS are discussed in Section 4. Finally, Section 5 draws some of the conclusions of the assessments carried out here.

## 2. MATHEMATICAL MODELING

In this section the main mathematical models accounted for in this work are briefly described. Governing equations are firstly discussed, followed by the sub-grid scale models employed here.

### 2.1. Governing equations

Compressible turbulent inert flows are considered in this study. Therefore, the Navier-Stokes equations plus the transport equations for mass and energy are solved here. In particular, following the LES-based approach, this set of equations is spatially filtered using a filtering operation of the form (Poinsot and Veynante, 2005),

$$\langle f(x,t) \rangle_l = \int_{-\infty}^{+\infty} f(x',t) F(x',x) dx', \tag{1}$$

where $\langle f(x,t) \rangle_l$ is the filtered value of $f(x',t)$ and $F(x',x)$ the filtered function. Additionally, to reduce filtered density-weighted terms, the Favre filtering is introduced as $\langle \rho \rangle_l \langle f(x,t) \rangle_L = \langle \rho f(x,t) \rangle_l$. After carrying out the LES filtering operation, the transport equations for mass, momentum, and energy accounted for here read as follows (Garnier et al., 2009),

$$\frac{\partial \langle \rho \rangle_l}{\partial t} + \frac{\partial \langle \rho \rangle_l \langle u_j \rangle_L}{\partial x_j} = 0, \tag{2}$$

$$\frac{\partial \langle \rho \rangle_l \langle u_i \rangle_L}{\partial t} + \frac{\partial \langle \rho \rangle_l \langle u_j \rangle_L \langle u_i \rangle_L}{\partial x_j} + \frac{\partial \langle p \rangle_l}{\partial x_i} - \frac{\partial \langle \sigma_{ij} \rangle_L}{\partial x_j} = -\frac{\partial \tau_{ij}}{\partial x_j}, \tag{3}$$

$$\frac{\partial \langle \rho \rangle_l \langle E \rangle_L}{\partial t} + \frac{\partial}{\partial x_j}\left[ (\langle \rho \rangle_l \langle E \rangle_L + \langle p \rangle_l) \langle u_j \rangle_L + \langle q_j \rangle_L - \langle \sigma_{ij} \rangle_L \langle u_i \rangle_L \right] = -\frac{\partial}{\partial x_j}\left( \gamma C_v Q_j + J_j - D_j \right). \tag{4}$$

In these last equations, $t$ and $x \equiv x_i$ ($i = 1,2,3$) are the temporal and spatial independent variables, respectively, $\rho$ is the density, $u$ represents velocity, $p$ pressure, and $E = C_v T + \frac{u_i u_i}{2}$ is the total energy. In addition, $C_v$ and $C_p$ are the heat capacity at constant volume and pressure, respectively, $T$ is the temperature, and $\gamma$ is the ratio of the specific heats. It is worth mentioning that density, pressure and temperature are linked through the ideal gas equation of state $p = \rho RT$. Moreover, $\sigma_{ij}$ and $q_j$ stand for the diffusive momentum and heat transport fluxes, respectively. The filtered form of these diffusive fluxes are given by,

$$\langle \sigma_{ij} \rangle_L = 2 \langle \mu \rangle_L \langle S_{ij} \rangle_L - \frac{2}{3} \langle \mu \rangle_L \delta_{ij} \langle S_{kk} \rangle_L, \quad \langle S_{ij} \rangle_L = \frac{1}{2}\left( \frac{\partial \langle u_i \rangle_L}{\partial x_j} + \frac{\partial \langle u_j \rangle_L}{\partial x_i} \right), \tag{5}$$

$$\langle q_j \rangle_L = -\langle \lambda \rangle_L \frac{\partial \langle T \rangle_L}{\partial x_i}, \tag{6}$$

where $\mu$ and $\lambda$ represent the dynamic viscosity and the thermal conductivity, respectively. Due to the LES filtering process, additional terms, i.e., the SGS stress $\tau_{ij}$, the SGS heat flux $Q_j$, the SGS turbulent diffusion $J_j$, and the SGS viscous diffusion $D_j$, appear in the right-hand side (RHS) of Eqs. (3) and (4). As expected, to solve the system of equations governing the flow of interest here, these SGS terms must be closed by an appropriate model.

**2.2. Closures for the SGS terms**

When closing the SGS additional terms present in LES-based approaches, the energy transfer mechanism from large to small scales that take place during turbulence dynamics is the main physical phenomenon accounted for (Garnier et al., 2009). To introduce this mechanism in which smallest scales end up dissipating the transferred energy, the Boussinesq hypothesis is adapted to a compressible framework by an equation of the form,

$$\tau_{ij} - \frac{1}{3} \delta_{ij} \tau_{kk} = -2 \langle \rho \rangle_l \nu_{sgs} \left( \langle S_{ij} \rangle_L - \frac{1}{3} \delta_{ij} \langle S_{kk} \rangle_L \right), \quad \tau_{kk} = 2 C_I \langle \rho \rangle_l \Delta^2 |\langle S_{ij} \rangle_L|^2, \tag{7}$$

where $\delta_{ij}$ stands for the Kronecker delta function, $\nu_{sgs}$ for the SGS viscosity, and $\tau_{kk}$ for the isotropic part of the SGS stress tensor computed by the Yoshizawa model. In addition, $C_I = 0.09$ is the Yoshizawa constant model, and $\Delta = \sqrt[3]{\Delta_x \Delta_y \Delta_z}$ is the filter size, which is a function of the grid spacing in each spatial direction. Moreover, to close the SGS stress tensor $\tau_{ij}$, the SGS viscosity is modeled. For the sake of argument, the details of the SGS viscosity modeling are postponed to Section 2.3. Following a similar construction to the one used in the SGS stress, the SGS heat flux $Q_j$ is assumed to be proportional to the resolved temperature gradient (Pino et al., 2000),

$$Q_j = -\frac{\langle\rho\rangle_l \nu_{sgs}}{Pr_t}\frac{\partial\langle T\rangle_L}{\partial x_i}, \tag{8}$$

where $Pr_t = 0.71$ is the turbulent Prandtl number. Notice from Eq. (8) that the SGS heat flux is directly linked to the SGS viscosity. Finally, the SGS turbulent diffusion is modeled by an expression of the form $J_j \cong \langle u_k\rangle_L \tau_{jk}$, whereas, due to its small influence in Eq. (4), the SGS viscous diffusion $D_j$ is neglected in this study.

### 2.3. Sub-grid scale models

SGS viscosity models attempt to mimic the drain of energy associated with the forward energy cascade (Pope, 2000). In this work, four different zero-equation SGS viscosity models are accounted for, (i) the classical Smagorinsky model, (ii) the dynamic Smagorinsky model, (iii) the wall-adapting local eddy-viscosity (WALE) model, and (iv) the Vreman model. The Smagorinsky model (Smagorinsky, 1963) is the simplest still-popular SGS model. In this model, the eddy viscosity is computed as,

$$\nu_{SGS} = (C_S\Delta)^2|\langle S_{ij}\rangle_L| = (C_S\Delta)^2\sqrt{2\langle S_{ij}\rangle_L:\langle S_{ij}\rangle_L}, \tag{9}$$

where $C_S = 0.16$ is the Smagorinsky model constant. Due to its dependency on the $C_S$ value, this model is generally over-dissipative and cannot reproduce the near wall behavior (Versteeg and Malalasekera, 2007). In the dynamic Smagorinsky model in turn, one of the variants of the classical Smagorinsky one, a dynamic procedure is carried out where the Smagorinsky constant $C_S$, the Yoshizawa one $C_I$, and the turbulent Prandtl number $Pr_T$ are spatially and temporally determined by expressions of the form (Pino et al., 2000),

$$C_S^2 = \frac{\langle\mathcal{L}_{ij}M_{ij}\rangle}{\langle M_{kl}M_{kl}\rangle}, \qquad C_I = \frac{\langle\mathcal{L}_{kk}\rangle}{\langle 2\{\Delta\}^2\{\langle\rho\rangle_l\}_l\{|\langle S_{ij}\rangle_L|^2\}_L - \{2\langle\rho\rangle_l\Delta^2|\langle S_{ij}\rangle_L|^2\}_l\rangle}, \qquad Pr_T = \frac{C_S^2\langle T_k T_k\rangle}{\langle\mathcal{K}_j T_j\rangle} \tag{10}$$

$$\mathcal{L}_{ij} = \left\{\frac{\langle\rho u_i\rangle_l\langle\rho u_j\rangle_l}{\langle\rho\rangle_l}\right\}_l - \frac{\{\langle\rho u_i\rangle_l\}_l\{\langle\rho u_j\rangle_l\}_l}{\{\langle\rho\rangle_l\}_l}, \qquad M_{ij} = \beta_{ij} - \{\alpha_{ij}\}_l,$$

$$\beta_{ij} = -2\{\Delta\}^2\{\langle\rho\rangle_l\}_l\left|\{\langle S_{ij}\rangle_L\}_L\right|\left(\{\langle S_{ij}\rangle_L\}_L - \frac{1}{3}\delta_{ij}\{\langle S_{kk}\rangle_L\}_L\right)$$

$$\alpha_{ij} = -2\langle\rho\rangle_l\Delta^2|\langle S_{ij}\rangle_L|\left(\langle S_{ij}\rangle_L - \frac{1}{3}\delta_{ij}\langle S_{kk}\rangle_L\right) \tag{11}$$

$$T_j = -\{\Delta\}^2\{\langle\rho\rangle_l\}_l\{|\langle S_{ij}\rangle_L|\}_L\frac{\partial\{\langle T\rangle_L\}_L}{\partial x_j} + \Delta^2\left\{\langle\rho\rangle_l|\langle S_{ij}\rangle_L|\frac{\partial\langle T\rangle_L}{\partial x_j}\right\}_l,$$

$$\mathcal{K}_j = \left\{\frac{\langle\rho u_j\rangle\langle\rho T\rangle_l}{\langle\rho\rangle_l}\right\}_l - \frac{\{\langle\rho u_j\rangle\}_l\{\langle\rho T\rangle_l\}_l}{\{\langle\rho\rangle_l\}_l}.$$

As it can be noticed from the above expressions, the dynamic procedure involves a test filter size $\{\Delta\} = 2\Delta$ at which transport quantities are constructed from the resolved ones obtained using the filter size $\Delta$. Moreover, the Fabre filtering at the test filter is defined as $\{\langle\rho\rangle_l\}_l\{\langle f(x,t)\rangle_L\}_L = \{\langle\rho f(x,t)\rangle_l\}_l$. Finally, in terms of simplicity, the WALE and Vreman models are similar to the classical Smagorinsky one. Indeed, the SGS viscosity in the WALE (Nicoud and Ducros, 1999) model is determined from,

$$\nu_{SGS} = (C_W\Delta)^2\frac{\left(S_{ij}^d:S_{ij}^d\right)^{\frac{3}{2}}}{\left(\langle S_{ij}\rangle_L:\langle S_{ij}\rangle_L\right)^{\frac{5}{2}} + \left(S_{ij}^d:S_{ij}^d\right)^{\frac{5}{4}}}, \tag{12}$$

$$S_{ij}^d = \frac{1}{2}\left(\langle g_{ij}\rangle_L^2 + \langle g_{ij}\rangle_L^2\right) - \frac{1}{3}\delta_{ij}\langle g_{kk}\rangle_L^2, \qquad \langle g_{ij}\rangle_L^2 = \langle g_{ik}\rangle_L\langle g_{kj}\rangle_L, \qquad \langle g_{ij}\rangle_L = \frac{\partial\langle u_i\rangle_L}{\partial x_j}, \tag{13}$$

where $C_W = 0.5$ is the WALE model constant, whereas the SGS viscosity model proposed by Vreman (2004) is of the form,

$$\nu_{SGS} = C \sqrt{\frac{B_\beta}{\alpha_{ij}:\alpha_{ij}}}, \tag{14}$$

$$\alpha_{ij} = \frac{\partial \langle u_i \rangle_L}{\partial x_j}, \quad \beta_{ij} = \Delta^2 \, \alpha_{mi} \, \alpha_{mj}, \quad B_\beta = \beta_{11}\beta_{22} - \beta_{12}^2 + \beta_{11}\beta_{33} - \beta_{13}^2 + \beta_{22}\beta_{33} - \beta_{23}^2, \tag{15}$$

where $C = 2.5 \, C_S^2$ is the Vreman model constant, which is a function of the Smagorinsky model one.

## 3. NUMERICAL APPROACH

In this section the mean features of the numerical approach employed here are briefly described.

### 3.1. Numerical methods

In the present work, the PeleC (PeleC, 2023) code was used to solve the set of governing equations described in Section 2. To handle the complex physics of compressible flows, the second-order predictor-corrector Godunov method and the hybrid PPM/WENO-Z one were established as the temporal and spatial schemes, respectively. Here, according to Motheau and Wakefield (2020), the purpose of the seventh-order WENO (weighted essentially non-oscillatory) scheme, including the Z version, is only to interpolate the primitive variables at cell faces in the PPM (piecewise parabolic method). Additionally, like Motheau and Wakefield (2020), all numerical simulations carried out here employed the gamma-law equation of state with $\gamma = 1.4$.

### 3.2. Computational domain and boundary conditions

Numerical simulations of the freely decaying version of the compressible HIT were performed in a $2\pi$ size cubic box (Figure 1) with periodic boundary conditions in all three spatial directions. Moreover, the initial velocity field was built from a prescribed energy spectrum given by,

$$E(k) \sim k^4 exp\left(-2\left(\frac{k}{k_0}\right)^2\right), \quad \frac{3u_{rms,0}^2}{2} = \frac{\langle u_0 \cdot u_0 \rangle}{2} = \int_0^\infty E(k)dk; \tag{16}$$

where $k_0 = 4$ is the most energetic wavenumber considered in this work, and $u_{rms,0}$ stands for the $u_{rms}$ in the initial condition, i.e., at $t = 0$. For comparison purposes with DNS data from Motheau and Wakefield (2021), the turbulent Mach and the Taylor Reynolds numbers were set at $M_{t,0} = 0.6$ and $Re_{\psi,0} = 100$, respectively. Notice that these two nondimensional parameters are related to the fluid viscosity through the following expressions,

$$M_{t,0} = \frac{\sqrt{\langle u_0 \cdot u_0 \rangle}}{c_0}, \tag{17}$$

$$Re_{\psi,0} = \frac{\rho_0 \psi_0 u_{rms,0}}{\mu_0}, \quad u_{rms,0} = \sqrt{\frac{\langle u_0 \cdot u_0 \rangle}{3}}, \quad \psi_0 = \frac{2}{k_0}, \tag{18}$$

where $c_0$ is the speed of sound at the beginning of the numerical simulations. As noticed from Eq. (18), the viscosity $\mu_0$ is computed from the Taylor Reynolds number expression. In addition, the thermal conductivity is defined by $\lambda_0 = \mu_0 C_p/Pr$ where $Pr$ is the Prandtl number, set to $Pr = 0.71$, and $C_p$ the specific heat capacity set equal to $C_p = 1.173 \, kJ/kgK$. Similar to the work done by Motheau and Wakefield (2021), the viscosity $\mu$ and the thermal conductivity $\lambda$ were held constant, i.e., $\mu = \mu_0$ and $\lambda = \lambda_0$. Finally, the initial pressure and temperature were established as being equal to $P_0 = 101.3 \, kPa$ and $T_0 = 1200 \, K$, respectively. All numerical simulations were carried out here accounting for a computational mesh featuring $256^3$ cells equally distributed in all three spatial directions, and with the same initial condition obtained when applying the inverse Fourier transform to Eq. (16). For numerical stability purposes, the timestep was set to be a variable and a function of the Courant-Friedrichs-Levy (CFL) number, which was kept constant at a value of 0.5. The maximum physical time considered here was $t/\tau = 4$ where $\tau = \psi_0/u_{rms,0}$.

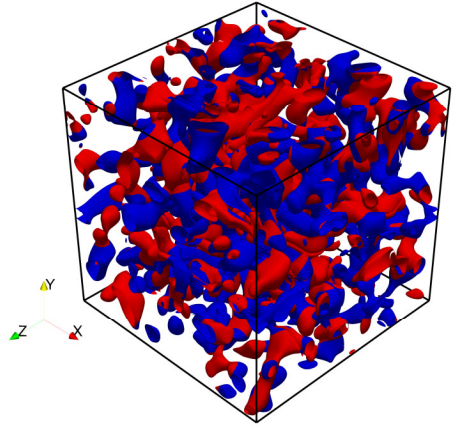

Figure 1. Iso-surfaces of Q-criterion $\frac{Q}{Q_{max}} = 0.1$ from the initial velocity field (blue and red colors indicate vorticity values of $\omega_z < 0$ and $\omega_z > 0$, respectively).

### 3.3. Mesh quality

Mesh quality is a critical aspect when carrying out LES numerical simulations of turbulent flows. In the past, to determine LES grid resolution requirements, several criteria have been proposed. These criteria include resolution ratios of the resolved to the total turbulent kinetic energy, energy spectra, and two-point correlations. From these criteria, it has been argued that two-point correlations are the best approach to ensure a well-resolved LES simulation (Wurps et al., 2020), so they have been used in this work (Figure 2). More specifically, grid resolution requirements have been determined here using a two-point correlation analysis of velocity components (Wurps et al., 2020),

$$B_{uu}(x^*) = \frac{\langle u'(x)u'(x-x^*)\rangle}{\sigma_u^2}, \quad (19)$$

where $u'$ and $\sigma^2$ stand for the u-velocity fluctuation and variance, respectively. Moreover, $x^*$ is the increasing distance between cells centers (points) over the line of interest at which information (i.e., velocity) is extracted from the solution domain. Therefore, when this distance $x^*$ is zero, $B_{uu}(0) = 1$ expressing the strong self-correlation of a point. Whereas, as the distance $x^*$ increases, $x^* \to \infty$, $B_{uu}(x^*)$ tends to a zero value. However, due to periodic boundary conditions, $B_{uu}(x^*)$ may not decrease to zero (Wurps et al., 2020). Following the analyses performed by Wurps et al. (2020) and Davidson (2009), it is assumed here that the quantity of cells describing a turbulent structure has values of $B_{uu}(x^*) > 0.3$. For the limiting case of $B_{uu}(x^*) = 0.3$, the distance $x^*$ is defined as $\sigma$, which represents the average resolved eddy size in the flow. In this case, the number of cells that describes the dominant resolved eddies can be obtained by $n_c = \sigma/\Delta$, where $\Delta$ is the cell size along the direction of the line of interest.

Accordingly, Figure 2.a shows the two-point correlations of the $u$-velocity component obtained accounting for four different mesh resolutions, i.e., $64^3$, $128^3$, $192^3$, and $256^3$. Notice that all length-related parameters, i.e., eddy dominant size, distance between points, and cell size, are normalized by $L$, which represent the half size of the cubic box. Similar to the analysis carried out by Wurps et al. (2020), the convergence of $B_{uu}(x^*)$ is evaluated i) in the upper part, i.e., $B_{uu}(x^*) \in [0.8; 1.0]$, ii) in the rapidly decrease to $B_{uu}(x^*) = 0.3$, and iii) when increasing the distance between points $B_{uu}(x^* \to \infty)$. As it can be noticed from Figure 2.a, the two-point correlation of the u-velocity component obtained with a $192^3$ mesh resolution is quite like the one obtained with a $256^3$ mesh resolution. In addition, Figure 2.b shows the size of the dominant resolved turbulent eddies associated with the four meshes assessed here, which highlights that the sizes of the dominant eddies described by the $192^3$ and $256^3$ meshes are almost the same. However, to capture the eddy dominant size with the largest mesh, about 10 cells more than for the case of a $192^3$ mesh resolution are required (Figure 2.c). Therefore, a mesh featuring $192^3$ elements seems to be enough to analyze the HIT discussed here. Nevertheless, as indicated above, to ensure the description of a well-resolved flow, a computational mesh featuring $256^3$ cells equally distributed in all three spatial directions has been used in all numerical simulations carried out in this work.

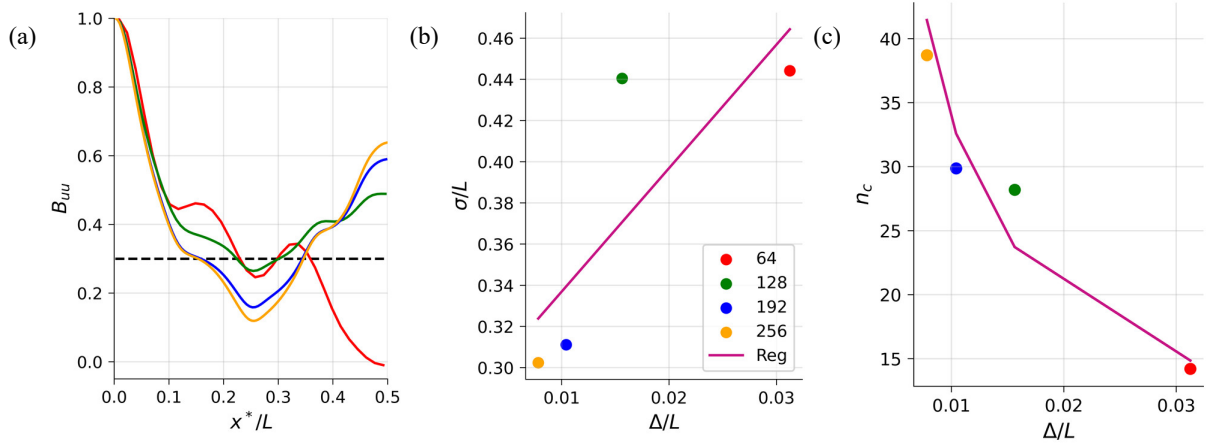

Figure 2. a) Normalized two-point correlation $B_{uu}$, b) Normalized size of the dominant resolved eddies, and c) number of cells describing the dominant resolved eddies. Lines in plots b and c represent values obtained from linear regression processes.

## 4.     RESULTS AND DISCUSSION

The main numerical results of the compressible HIT in its freely decaying version obtained here are presented and discussed in this section.

### 4.1. Decay over time results

As the initial energy associated with the flow field is not maintained in this work, physical quantities such as kinetic energy, enstrophy, temperature variance, and dilatation decrease over time. This aspect is clearly observed in Figure 3, which shows the temporal evolution of these quantities for each of the SGS models accounted for here. Notice that for comparison purposes with the DNS data from Motheau and Wakefield (2020), both physical quantities and time values are shown in nondimensional terms. Additionally, due to the isotropic and homogeneous properties of the flow, all cells at any given time can be treated in the same way, improving thus the sample size. In Figure 3, the angle brackets $\langle \rangle$ that appears in the left axis of each subplot represent the cell volume weighted average of a particular quantity. In accordance with the DNS data (black solid line), the numerical results of the flow physical quantities studied here indicate that these parameters are significantly influenced by the strong compressibility effects of the flow (Motheau and Wakefield, 2020). Unlike the kinetic energy (Figure 3.a) indeed, which shows only a slight rise along its decreasing behavior around time $t/\tau = 0.25$, the other analyzed quantities exhibit a piecewise monotonic behavior. In particular, as noticed from Figure 3, the enstrophy, temperature variance, and dilatation present a peak value at times $t/\tau = 1.5, 0.5, 0.75$, respectively, after which the initially increasing behavior switches to a decreasing one.

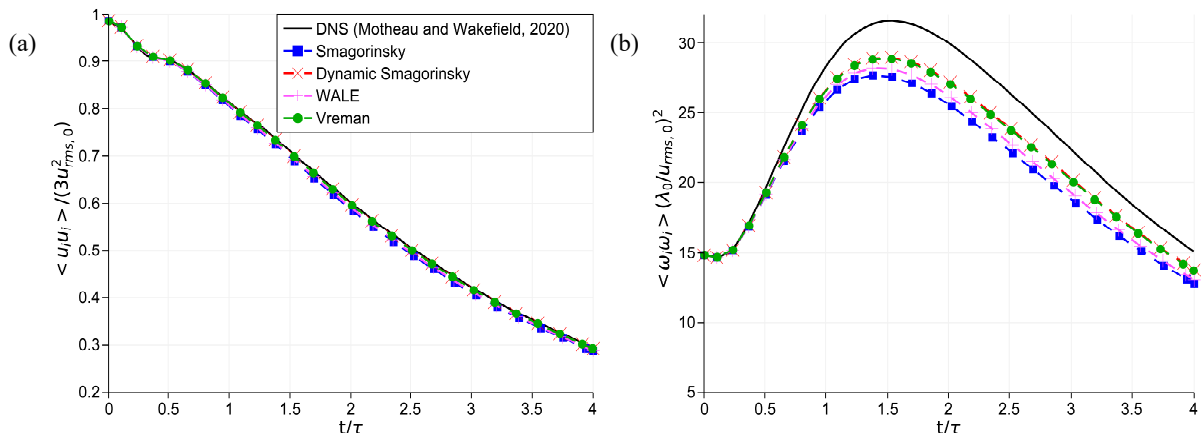

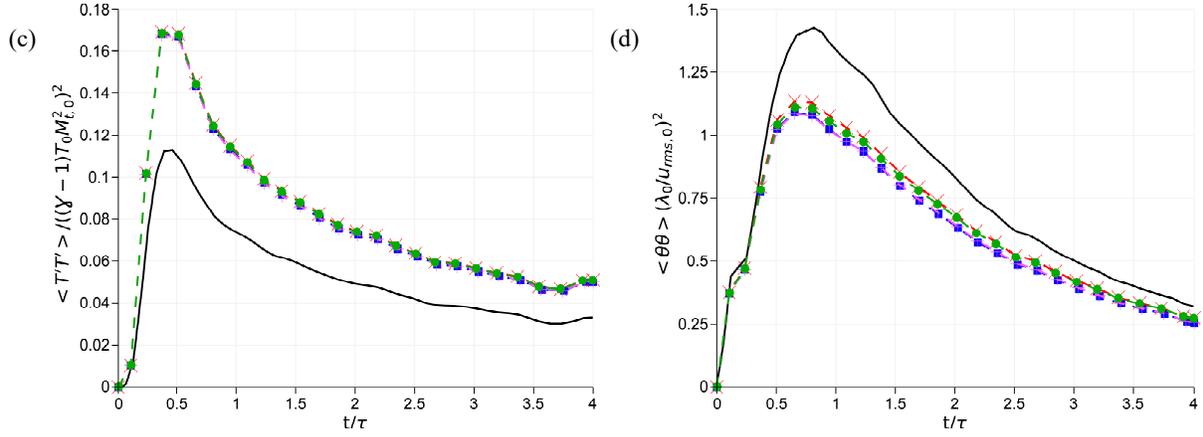

Figure 3. Temporal evolution of different physical quantities. a) kinetic energy, b) enstrophy, c) temperature variance, and d) dilatation.

The numerical results obtained here show particularly that all SGS models are capable to describe the DNS kinetic energy decay (Figure 3.a). In addition, even though the SGS models also capture the enstrophy and dilatation expected trends qualitatively, the LES predicted values are underestimated. Thus, compared to the corresponding DNS results, the peak values of enstrophy (Figure 3.b) and dilatation (Figure 3.d) are underestimated by about 9.1% and 18.5%, respectively. Contrarily, the peak value of the temperature variance (Figure 3.c) is overestimated by about 53.4%. It is also observed from Figure 3.b that the dynamic Smagorinsky and Vreman model results are overlapped and closer to the DNS data than those results obtained by the WALE and Smagorinsky models. Similarly, in Figure 3.d, the WALE and the Smagorinsky model results are overlapped and farther to the DNS results than the ones obtained using the Vreman and dynamic Smagorinsky models.

### 4.2. Turbulent spectra results

One of the of the main complexities of compressible flows is that there is more than one energy transfer mechanism (Gatski and Bonnet, 2013). Unlike the incompressible case, energy is not purely transferred between turbulent scales, but also between modes (Garnier et al., 2009). According to Kovasznay (1953), compressible turbulent fluctuations can be viewed as a combination of three fundamental modes, (i) a vorticity mode, (ii) an acoustic (pressure) mode, and (iii) an entropy mode. The different energy transferred paths are illustrated in Figure 4, where the spectra of selected physical quantities such as kinetic energy, vorticity, dilatation and density are shown. It is worth mentioning here that each of these physical quantities are linked to the inter-scale and intermodal energy transfer (Garnier et al., 2009, Gatski and Bonnet, 2013). For verification purposes, DNS results provided by Motheau and Wakefield (2020) are included in this figure as well. As it can be observed from Figure 4, the four SGS models studied here properly describe the DNS spectra trends, and the small discrepancies observed appear only around wave numbers $k = 48$ and $64$ for vorticity (Figure 4.b) and dilatation (Figure 4.c) spectra, respectively. Overall, the numerical results obtained with the SGS models accounted for here in relatively high wave numbers do not present a pile-up of physical quantities as the ones reported in the work with no LES approach by Motheau and Wakefield (2020).

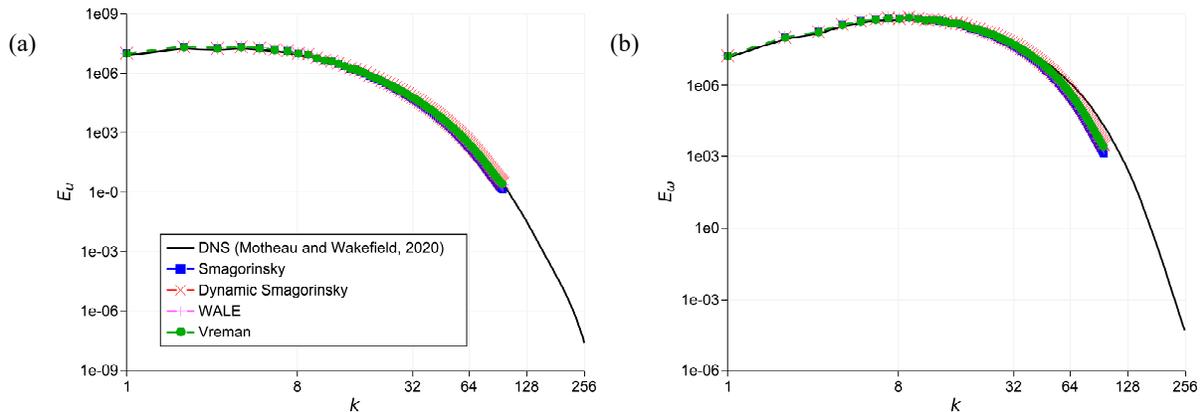

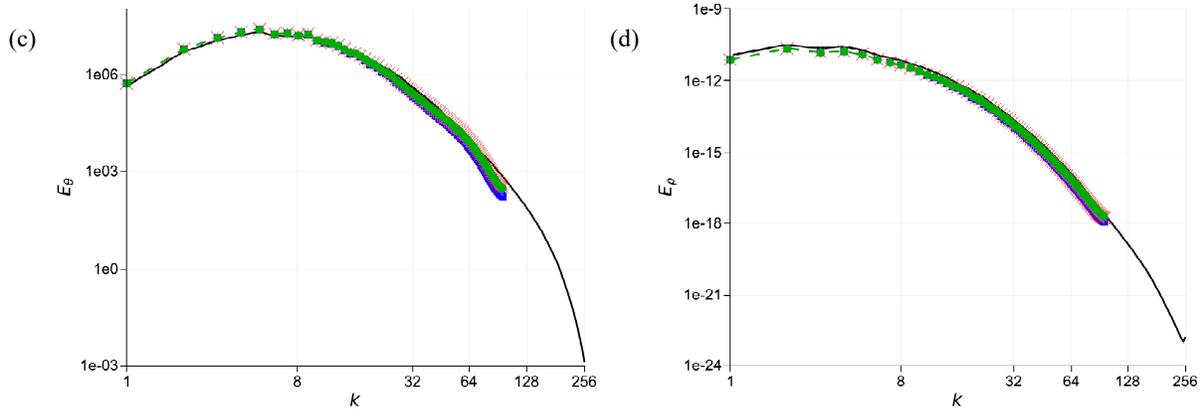

Figure 4. Spectra of different physical quantities at $t/\tau = 4$. a) kinetic energy, b) vorticity, c) dilatation, and d) density.

## 5. CONCLUSIONS

In this work, four classical LES SGS models, being two of them implemented into PeleC, have been evaluated for the freely decaying version of compressible HIT. More specifically, to have an insight of the flow characteristics, the temporal evolution of some of the main physical quantities characterizing it were compared with DNS results available in literature. From the LES predicted solutions, it was found that the SGS models considered here are capable of capturing the overall trends of all physical quantities, but only the predicted kinetic energy closely matches the corresponding DNS data. In particular, it was found that, compared to the Smagorinsky and WALE models, the dynamic Smagorinsky and Vreman ones provide better approximations to the DNS data. However, in comparison to the Vreman model, the dynamic Smagorinsky model is more computationally expensive. Although the associated results were not included in this work, it was also observed that refining the computational mesh ($64^3$, $128^3$, and $256^3$) improves the enstrophy predictions. Indeed, in the DNS work used as reference here, a mesh size of $512^3$ and no SGS modeling was used in the simulations carried out there and the results agreed relatively well with the DNS data. Of course, the mesh size used in the referred work contradicts the original purpose of LES that aims to provide accurate predictions with fewer computational resources compared to DNS. Finally, the discrepancies between LES and DNS results observed in this work may be attributed to the limitations of both the Yoshizawa model and the SGS models studied here to accurately describe turbulent compressible flows. Overall, the results obtained in this work highlight the need of improved SGS models capable of adequately describing turbulence dynamics in compressible flows.

## 6. ACKNOWLEDGEMENTS

This work has been supported by the US Army Research Laboratory under Research Grant No. W911NF-22-1-0275. Luis Bravo was supported by the US Army Research Laboratory 6.1 Basic research program in propulsion sciences.

## 8. RESPONSIBILITY NOTICE